\documentclass[11pt]{article}
\usepackage{amssymb}
\usepackage{amsfonts}
\usepackage{amsmath}
\usepackage{geometry}
\usepackage{graphicx,booktabs,longtable,array,calc,pdfpages}
\usepackage[T1]{fontenc}
\usepackage[utf8]{inputenc}
\usepackage{microtype}
\usepackage{hyperref}

\newtheorem{theorem}{Theorem}
\newtheorem{proposition}{Proposition}
\newtheorem{assumption}{Assumption}
\newtheorem{condition}{Condition}
\newtheorem{remark}{Remark}
\newenvironment{proof}[1][Proof]{\noindent\textbf{#1.} }{\ \rule{0.5em}{0.5em}}

\begin{document}

\title{Discrete Choice under Congestion: Equilibrium and Welfare}
\author{Debopam Bhattacharya \\
University of Cambridge \and Matias Makiranta \\
University of Helsinki}
\date{29 September 2026}
\maketitle

\textbf{Abstract}: In many choice settings, a rise in aggregate demand for
an option lowers an individual's payoff from choosing it. Driving becomes
less attractive than taking a train when more commuters drive, clogging the
highway. In such cases, a targeted intervention that mitigates the
externality at the congested option can divert it elsewhere and also
undermine preference based sorting, making overall welfare effects
theoretically ambiguous. We develop an empirical framework to analyze such
scenarios of endogenous congestion. We show that in these settings,
equilibrium exists and is unique without restrictions on functional-form of
utilities or dimension/distribution of preference heterogeneity, and even
when congestion at one option affects payoffs from other choices. Under a
linear-index structure, micro-data on choice enables semiparametric
identification of utility contrasts and welfare effects of some, but not
all, interventions. We apply the framework to Cambridge undergraduate
admissions, where candidates apply through one constituent college, and
admission probability depends on application numbers relative to capacity in
that college. Using administrative micro-data on 25,325 applicants, we
calculate welfare under two counterfactuals: (i) centralized random
allocation and (ii) higher rent at popular colleges. Random allocation
reduces welfare for everyone: it removes individual choice without improving
admission chances since, by design, lower congestion at a college exactly
counterbalances the lower probability of being allocated there. Targeted
rent increase produces both gains and losses, with median loss for men and
median gain for women. This exercise shows how identified preferences and
equilibrium feedback jointly determine welfare when choices entail
externalities.\pagebreak 

\section{Introduction}

A fundamental question in public economics is whether centralized assignment
systems should replace decentralized choice when individual sorting creates
negative externalities. In settings ranging from higher education to urban
transport, a surge in aggregate demand for an option often lowers its payoff
to individuals who choose it. While policy interventions can mitigate these
congestion externalities, they simultaneously distort individual
preference-based sorting. Because these two forces operate in opposite
directions, the aggregate and distributional welfare effects of moving away
from free decentralized choice remain theoretically ambiguous. Assessing
them, therefore, requires an empirical framework that maps heterogeneous
micro-preferences under negative congestion externalities into unique
policy-relevant equilibria.

In this paper, we develop a comprehensive empirical framework for studying
discrete choice when an alternative's payoff depends endogenously on
aggregate choice shares. Decision-makers optimally sort across options while
anticipating these aggregate shares; the aggregate shares, in turn, are
determined by the joint distribution of individual choices. Evaluating
counterfactual policies in such environments introduces three challenges:
equilibrium multiplicity, structural unidentifiability of welfare levels,
and the confounding presence of behavioral biases in probability perception.
We resolve these challenges theoretically and provide corresponding
empirical tools for measuring welfare effects in real-life settings with
congestion, using cross-sectional microdata.

Our first and core theoretical contribution is a non-parametric unique
equilibrium theorem under a condition we term \textit{relative congestion
dominance}. In economic environments characterized by negative
externalities, policy interventions frequently generate multiple
counterfactual choice configurations, rendering welfare predictions
indeterminate even if all structural parameters are known. We prove that if
an environment satisfies relative congestion dominance---meaning that when
congestion shifts toward a set of alternatives, each member of that set
becomes weakly less attractive relative to every alternative outside
it---the equilibrium choice vector exists and is unique. Crucially, our
proof bypasses restrictive functional forms, additive separability, linear
interaction indices, extreme-value error distributions, or smooth
contraction mappings that standard frameworks maintain. It demonstrates the
fundamental link between negative externalities and equilibrium determinacy,
thereby expanding the boundaries of counterfactual policy simulation in
discrete choice settings.

Our second contribution addresses the empirical identification of
counterfactual welfare distributions. We establish that when utilities have
a linear index structure with an unknown error distribution, multinomial
index inversion can identify systematic utility contrasts. By exploiting
repeated choice environments, we isolate common option components from
unobserved individual shocks. Importantly, we demonstrate a sharp contrast
in policy identifiability: while the welfare distributions of centralized
random allocations and congestion charges are point-identified directly from
choice data, the welfare impacts of altering individual covariates remain
fundamentally unidentified without ad-hoc functional extrapolations.

Our third contribution integrates behavioral decision-making into the core
framework of choice under congestion. In high-stakes sorting environments,
individuals may not possess perfect rational expectations, and instead
systematically distort objective success probabilities. By embedding
behavioral probability weighting into our discrete choice framework, we
document how subjective distortions interact with equilibrium feedback. For
the sinle-index setting, we prove that variation in objective success
probabilities across distinct markets or choices identifies both the
underlying structural preference parameters and the behavioral probability
distortion function. This allows us to disentangle two distinct sources of
welfare loss: the technological externality driven by physical capacity
constraints, and the behavioral externality driven by misperceived
probabilities. Consequently, we can evaluate a different class of policies,
viz. information interventions that correct probability misperceptions
without distorting markets.

We demonstrate the empirical power of our framework by applying it to
undergraduate admissions at the University of Cambridge. This setting
provides an ideal context for our purpose: applicants are permitted
completely free decentralized choice among constituent colleges, colleges
have identical preference for students dictated by central
university/department criteria, but each college faces strict capacity
constraints set in advance, meaning that a higher application share at a
specific college lowers an applicant's ex ante probability of receiving an
offer. Utilizing administrative micro-data on 25,325 applicants, we estimate
gender-specific structural preference parameters and solve for the global
fixed points under alternative market interventions.

Our empirical results yield insights that contrast sharply with conventional
policy intuitions. First, we show that transitioning to centralized random
allocation generates universal welfare losses for both men and women.
Essentially, the systemic reduction in congestion at popular colleges is
perfectly counterbalanced by a lower probability of being allocated there,
leaving applicants with identical acceptance probabilities as in the
decentralized case while eroding the option value of choice. Second, we
simulate a market-based congestion charge---imposing a rent premium at
oversubscribed colleges. This policy triggers complex equilibrium
substitutions across the entire college system, generating highly unequal
distributional outcomes: a median welfare \textit{loss} of \pounds 98.27 for
men, but a median welfare \textit{gain} of \pounds 27.59 for
women.\smallskip 

The rest of this paper is organized as follows. Section 2 develops the
general model, states and proves the uniqueness theorem, and defines the
interventions and their welfare effects. Section 3 describes Cambridge
admission process and its mapping to the model. Section 4 gives the
identification argument and clarifies which welfare counterfactuals are
identified directly from utility contrasts. Section 5 describes the data and
reports empirical results. Section 6 discusses behavioral choice, and
Section 7 concludes with suggestions for future work. All tables and figures
and a relevant extract from Cambridge's official admission statistics
publication appear at the end of the paper. Additional proofs and examples
are collected in an Appendix at the end.

\section{Set-up and Congestion Mechanism}

Consider a large population of individuals $i=1,\ldots ,I$, each choosing
one of $J$ alternatives. Let $\mathbf{t}_{ij}$ collect all observed and
unobserved determinants of individual $i$'s payoff from alternative $j$. If
the individual expects the vector $\boldsymbol{\pi }_{i}^{e}$ of population
choice shares, utility is 
\begin{equation}
\widetilde{U}_{j}(\mathbf{t}_{ij},\boldsymbol{\pi }_{i}^{e}).
\label{eq:utility}
\end{equation}%
At this stage we do not restrict the dimension of $\mathbf{t}_{ij}$, how its
components enter utility, whether they are observable to us or their joint
distribution.

\begin{assumption}
Individuals have rational expectations: $\boldsymbol{\pi }_{i}^{e}=%
\boldsymbol{\pi }$ for all $i$, where $\pi _{j}$ is the population fraction
choosing $j$. Utilities satisfy \emph{relative congestion dominance}: for
any two share vectors $\boldsymbol{\pi }$ and $\boldsymbol{\rho }$, let 
\begin{equation*}
A(\boldsymbol{\pi },\boldsymbol{\rho })=\{j:\pi _{j}\leq \rho _{j}\}.
\end{equation*}%
For every $a\in A(\boldsymbol{\pi },\boldsymbol{\rho })$ and $b\notin A(%
\boldsymbol{\pi },\boldsymbol{\rho })$, 
\begin{equation}
\widetilde{U}_{a}(\mathbf{t}_{ia},\boldsymbol{\rho })-\widetilde{U}_{b}(%
\mathbf{t}_{ib},\boldsymbol{\rho })\leq \widetilde{U}_{a}(\mathbf{t}_{ia},%
\boldsymbol{\pi })-\widetilde{U}_{b}(\mathbf{t}_{ib},\boldsymbol{\pi }).
\label{eq:relativecongestion}
\end{equation}%
\label{ass:negative}
\end{assumption}

Assumption \ref{ass:negative} restricts utility contrasts, and allows
congestion to have direct, heterogeneous spillover effects. The restriction
says that when congestion shifts toward a set of alternatives, every
alternative in that set becomes weakly less attractive relative to every
alternative outside it. As a specific example, suppose%
\begin{equation*}
\widetilde{U}_{j}(\mathbf{t}_{ij},\boldsymbol{\pi })=v_{ij}-\sum_{\ell
=1}^{J}c_{j\ell }\pi _{\ell },\qquad c_{j\ell }\geq 0.
\end{equation*}%
All congestion shares may then affect every alternative, and the
coefficients $c_{j\ell }$ can vary by $j,l$. Then condition (\ref%
{eq:relativecongestion}) becomes 
\begin{equation*}
c_{ar}-c_{br}\geq c_{as}-c_{bs}\quad \text{for }a,r\in A,\ b,s\notin A.
\end{equation*}%
Thus spillovers are allowed; what matters is that congestion shifted into a
set harms that set's relative attractiveness most.

\begin{remark}
A special case of the above assumption is where utility from an alternative
does not depend on congestion at other alternatives, holding own congestion
fixed, i.e., $\widetilde{U}_{j}(\mathbf{t}_{ij},\boldsymbol{\pi })\equiv 
\widetilde{U}_{j}(\mathbf{t}_{ij},\pi _{j})$ for all $j=1,...,J$ and $%
\widetilde{U}_{j}(\mathbf{t}_{ij},\pi _{j})$ is weakly decreasing in $\pi
_{j}$, which will imply (\ref{eq:relativecongestion}). This is the situation
in our empirical application, and as is assumed in Brock and Durlauf's
seminal theory papers, cited above. In this case. However, this special case
rules out some real life situations like traffic on one route delaying users
of other routes that share a road segment, which is covered by our more
general assumption \ref{ass:negative}.
\end{remark}

\begin{remark}
Note that direct cross-congestion and equilibrium substitution are distinct.
Direct cross-congestion occurs when the share choosing $k$ enters the 
\textbf{ex-ante} payoff from choosing $j$, so that $\frac{\partial U_{ij}}{%
\partial \pi _{k}}\neq 0$ for $k\neq j$. Equilibrium substitution can change
the ex-post share and payoff of $j$ even when this derivative is zero:
greater congestion at $k$ lowers the attractiveness of $k$, redirects agents
toward $j$, and thereby changes $\pi _{j}$. Thus the special case described
in Remark 1 still allows for spillover effects of congestion. It specifies
that \textbf{holding congestion at alternative }$j$\textbf{\ fixed},
congestion at other alternative $k\neq j$ do not affect the ex-ante utility
from choosing alternative $j$. As clarified in Remark 2, equilibrium
substitution resulting from increased congestion at $j$ can increase
congestion at $k$ and thereby reduce the ex-post utility of choosing $k$%
.\smallskip
\end{remark}

\textbf{Choice-probabilities and Equilibrium: }Given a candidate vector $%
\boldsymbol{\pi }=(\pi _{1},\ldots ,\pi _{J})$, define the choice
probabilities as%
\begin{equation}
G_{j}(\boldsymbol{\pi })={}\Pr \left[ \widetilde{U}_{j}(\mathbf{t}_{ij},%
\boldsymbol{\pi })>\max_{k\neq j}\widetilde{U}_{k}(\mathbf{t}_{ik},%
\boldsymbol{\pi })\right] .  \label{eq:G}
\end{equation}

We now state a final technical assumption, and prove a theorem that
establishes existence and uniqueness of equilibrium in this set-up. The
theorem does more than rationalize the observed equilibrium. It makes the
post-intervention allocation unique. Without it, a policy could generate
several counterfactual choice vectors and \textit{a set} of welfare effects
even if all preference parameters were known.\smallskip

\begin{assumption}
The vectors $(\mathbf{t}_{i1},\ldots ,\mathbf{t}_{iJ})$ are identically
distributed across $i$. The functions $\widetilde{U}_{j}$ and $G_{j}$ are
continuous in their arguments, and $\widetilde{U}_{j}(\mathbf{t}_{ij},%
\boldsymbol{\pi })-\widetilde{U}_{k}(\mathbf{t}_{ik},\boldsymbol{\pi })$ are
continuously distributed, i.e. for every $j\neq k$ and every $\boldsymbol{%
\pi }$,%
\begin{equation}
\Pr [\widetilde{U}_{j}(\mathbf{t}_{ij},\boldsymbol{\pi })=\widetilde{U}_{k}(%
\mathbf{t}_{ik},\boldsymbol{\pi })]=0.
\end{equation}%
\label{ass:ties}
\end{assumption}

The last condition is the usual no-ties restriction in discrete choice. It
is the only substantive restriction we impose on the distribution of
preference heterogeneity.

An equilibrium is a fixed point satisfying%
\begin{equation}
\boldsymbol{\pi }=G(\boldsymbol{\pi }),\qquad \sum_{j=1}^{J}\pi _{j}=1.
\label{eq:fixedpoint}
\end{equation}

\begin{theorem}
Under Assumptions \ref{ass:negative}--\ref{ass:ties}, the fixed-point
condition (\ref{eq:fixedpoint}) has a unique solution. \label{thm:unique}
\end{theorem}

\begin{proof}
Continuity of $G$ and the fact that it maps the unit simplex into itself
imply existence by Brouwer's fixed-point theorem.

For uniqueness, suppose instead that $\boldsymbol{\pi }\neq \boldsymbol{\rho 
}$ are two fixed points. Since both vectors sum to one, there is a $K$, $%
1\leq K<J$, such that 
\begin{equation}
\pi _{j}\leq \rho _{j},\quad j=1,\ldots ,K,\qquad \pi _{j}\geq \rho
_{j},\quad j=K+1,\ldots ,J,  \label{eq:partition}
\end{equation}%
with a strict inequality in each group for at least one alternative. Let $%
A=\{1,\ldots ,K\}$.

Suppose, if possible, that 
\begin{equation}
\sum_{j\in A}G_{j}(\boldsymbol{\pi })<\sum_{j\in A}G_{j}(\boldsymbol{\rho }).
\label{eq:massswitch}
\end{equation}%
Then a positive mass of individuals must choose an alternative in $A$ under $%
\boldsymbol{\rho }$ but an alternative outside $A$ under $\boldsymbol{\pi }$%
. Consider one such individual and denote the two alternatives by $a\in A$
and $b\notin A$. Her choices imply 
\begin{align}
\widetilde{U}_{a}(\mathbf{t}_{ia},\boldsymbol{\rho })& \geq \widetilde{U}%
_{b}(\mathbf{t}_{ib},\boldsymbol{\rho }),  \label{eq:rp1} \\
\widetilde{U}_{a}(\mathbf{t}_{ia},\boldsymbol{\pi })& \leq \widetilde{U}_{b}(%
\mathbf{t}_{ib},\boldsymbol{\pi }).  \label{eq:rp2}
\end{align}%
But (\ref{eq:partition}) and Assumption \ref{ass:negative} imply (see eqn. (%
\ref{eq:relativecongestion}))%
\begin{equation*}
\widetilde{U}_{a}(\mathbf{t}_{ia},\boldsymbol{\rho })-\widetilde{U}_{b}(%
\mathbf{t}_{ib},\boldsymbol{\rho })\leq \widetilde{U}_{a}(\mathbf{t}_{ia},%
\boldsymbol{\pi })-\widetilde{U}_{b}(\mathbf{t}_{ib},\boldsymbol{\pi }).
\end{equation*}%
Together with (\ref{eq:rp1})--(\ref{eq:rp2}), this forces both differences
to equal zero, contradicting Assumption \ref{ass:ties}. Hence (\ref%
{eq:massswitch}) is impossible and 
\begin{equation}
\sum_{j\in A}G_{j}(\boldsymbol{\pi })\geq \sum_{j\in A}G_{j}(\boldsymbol{%
\rho }).
\end{equation}%
Since both vectors are fixed points, 
\begin{equation}
\sum_{j\in A}\pi _{j}\geq \sum_{j\in A}\rho _{j},
\end{equation}%
which contradicts $\pi _{j}\leq \rho _{j}$ for every $j\in A$ with at least
one strict inequality. Therefore the equilibrium is unique.
\end{proof}

The intuition behind the proof is a revealed-preference argument. Moving
from one candidate equilibrium to another makes every alternative whose
share rises weakly less attractive \emph{relative to} every alternative
whose share falls. So aggregate demand cannot move toward the former set.
The argument allows congestion to affect all utilities and uses no
contraction, smoothness, or distributional restriction.\smallskip

\textbf{Connection with related results}: Our Theorem 1 is related to
Milchtaich's (2000) aggregate uniqueness result for large crowding games
where the special case described in Remark 1 above holds, viz. $\widetilde{U}%
_{j}(\mathbf{t}_{ij},\boldsymbol{\pi })\equiv \widetilde{U}_{j}(\mathbf{t}%
_{ij},\pi _{j})$ with $\widetilde{U}_{j}(\mathbf{t}_{ij},\pi _{j})$ \textit{%
strictly decreasing} in $\pi _{j}$, and also permits representative players
to randomize across tied best responses. In contrast to the set-up of
Milchtaich, we (i) require primitive utilities to have heterogeneous
spillover effects, i.e. utility for $j$ can depend on all of $\mathbf{\pi }$
and not just $\pi _{j}$, and (ii) require only \textit{weak} monotonicity of
payoffs in own congestion. These generalizations are empirically relevant
and significantly expand the scope of the uniqueness result. For example, in
one's choice between bus, car or train, excessive bus traffic hurts payoff
from driving if cars and buses share highway space but does not affect
utility from train. On the other hand, one's utility from driving need not
strictly decrease if one more person drives, unless it necessarily delays
one for work, i.e. congestion only matters when the capacity constraint
binds. In contrast, we impose the standard no-ties condition of discrete
choice models which is not used in Milchtaich. The two results are therefore
both mathematically and substantively different.

Theorem 1 is also reminiscent of Berry et al 2013's result on invertibility
of demand in multinomial choice settings under a `weak substitute' and a
`connected strict substitute' assumptions. Consider the special case,
described in Remark 1 above, where $\widetilde{U}_{j}(\mathbf{t}_{ij},%
\boldsymbol{\pi })\equiv \widetilde{U}_{j}(\mathbf{t}_{ij},\pi _{j})$ for
all $j=1,...,J$, and $\widetilde{U}_{j}(\mathbf{t}_{ij},\pi _{j})$ is weakly
decreasing in $\pi _{j}$; let $\sigma \left( \mathbf{\pi }\right) \equiv 
\mathbf{\pi }-G\left( \mathbf{\pi }\right) $. This implies that the "weak
substitute" assumption of Berry et al is satisfied by $\sigma \left( \cdot
\right) $ because $\sigma _{j}\left( \mathbf{\pi }\right) \equiv \pi
_{j}-G_{j}\left( \mathbf{\pi }\right) $ and $G_{j}\left( \mathbf{\pi }%
\right) ={}\Pr \left[ \widetilde{U}_{j}(\mathbf{t}_{ij},\pi
_{j})>\max_{k\neq j}\widetilde{U}_{k}(\mathbf{t}_{ik},\pi _{k})\right] $ is
weakly increasing in $\pi _{k}$ for $k\neq j$. However, their "connected
strict substitute" assumption need not hold for $\sigma \left( \cdot \right) 
$. To see this, note by Lemma 1 of Berry et al 2013 that for the connected
strict substitute assumption to hold, it is necessary that $\pi _{l}-$ $%
G_{l}\left( \mathbf{\pi }\right) $ be \textbf{strictly} decreasing in some $%
\pi _{k}$ with $k\neq l$, (i.e. set $\mathcal{K=}\left\{ 1,...,J\right\}
\backslash \left\{ l\right\} $ in Lemma 1 of Berry et al 2013). In contrast,
our assumption 1 only implies that $\pi _{l}-$ $G_{l}\left( \mathbf{\pi }%
\right) $ is \textbf{weakly} decreasing in \textit{each} $\pi _{k}$ with $%
k\neq l$.\footnote{%
This continues to hold even if we strengthen our assumption 1 to require
that $\tilde{U}_{j}\left( \mathbf{t}_{ij},\pi _{j}\right) $ be \textit{%
strictly} decreasing in $\pi _{j}$, unless we impose additional,
economically unjustified assumptions on the joint distribution of $\tilde{U}%
_{j}\left( \mathbf{t}_{ij},\pi _{j}\right) $'s and/or how unobserved
heterogeneity enters the $\tilde{U}_{j}\left( \mathbf{t}_{ij},\pi
_{j}\right) $'s.} On the other hand, our condition that $G_{j}\left( \mathbf{%
\pi }\right) $ is weakly decreasing in $\pi _{j}$, which is important for
the proof above, translates to: $\pi _{j}-\sigma _{j}\left( \mathbf{\pi }%
\right) $ is weakly decreasing in $\pi _{j}$, which has no analog among the
Berry et al assumptions. This, together with the counter-examples provided
in the appendix below, clarifies that our conditions and result on
uniqueness of fixed point are distinct from those of Berry et al on
invertibility. Furthermore, $\widetilde{U}_{j}(\mathbf{t}_{ij},\boldsymbol{%
\pi })\equiv \widetilde{U}_{j}(\mathbf{t}_{ij},\pi _{j})$ is only a special
case of our assumption \ref{ass:negative}.

Note that our theorem is also stronger than a standard contraction argument
which would require an upper bound on the response of choice probabilities
to congestion. Theorem \ref{thm:unique} restricts only the ordering of
relative responses. Congestion can have arbitrarily large effects, and those
effects can be heterogeneous and nonsmooth at the individual level, as long
as the aggregate choice map remains continuous.

\subsection{Interventions and their Welfare Effects}

We study two interventions that mitigate congestion while also distorting
choice.

\subsubsection{Centralized Allocation}

Under centralized random allocation, individuals are assigned to
alternatives in proportions determined by capacity. The policy can
potentially reduce congestion at initially popular alternatives, but it
removes the option value of being able to choose the preferred alternative.
The welfare effect of centralization can be calculated as follows. Let
decentralized equilibrium indirect utility for individual $i$ be $%
V_{i}^{D}=\max_{j}\widetilde{U}_{j}(\mathbf{t}_{ij},\boldsymbol{\pi })$.
Suppose centralized allocation assigns an individual to alternative $j$ with
probability $n_{j}/n$, where $n_{j}$ is capacity of $j$th option and $%
n=\sum_{j=1}^{J}n_{j}$. Expected utility is%
\begin{equation}
V_{i}^{C}=\sum_{j=1}^{J}\frac{n_{j}}{n}\widetilde{U}_{j}\left( \mathbf{t}%
_{ij},\mathbf{n}/n\right) .
\end{equation}%
where $\mathbf{n}/n$ denotes the vector $\left( \frac{n_{1}}{n},...,\frac{%
n_{J}}{n}\right) $.

The welfare effect of random allocation can be entirely nonparametrically
identified if one can conduct an appropriate counterfactual choice
experiment. To see this, suppose $\mathbf{t}_{ij}\equiv \left(
S_{ij},R_{j}\right) $ where $R_{j}$ is the numeraire and $\widetilde{U}%
_{j}\left( S_{ij},R_{j},\pi _{j}\right) $ is strictly increasing in $R_{j}$.
Then welfare effect of random allocation can be measured by compensating
variation $T_{i}$ which solves 
\begin{equation}
\max_{j}\widetilde{U}_{j}\left( S_{ij},R_{j},\boldsymbol{\pi }\right)
=\sum_{j=1}^{J}\frac{n_{j}}{n}\widetilde{U}_{j}\left( S_{ij},R_{j}+T_{i},%
\mathbf{n}/n\right) .
\end{equation}%
Because the right-hand side is strictly increasing in $T_{i}$, for any $\tau 
$,%
\begin{equation*}
\Pr \left( T_{i}\leq \tau \right) =\Pr \left( \sum_{j=1}^{J}\frac{n_{j}}{n}%
\widetilde{U}_{j}\left( S_{ij},R_{j}+T_{i},\mathbf{n}/n\right) \leq \tau
\right) =\Pr \left( \max_{j}\widetilde{U}_{j}\left( S_{ij},R_{j},\boldsymbol{%
\pi }\right) \leq \tau \right) \text{.}
\end{equation*}%
Hence the CDF of $T_{i}$ could, in principle, be recovered directly from an
experiment that offers individuals the above choice at different values of $%
\tau $. Although we do not have data from such an experiment and so this
route is not feasible for us, the above argument clarifies that
identifiability of welfare effects may not require any functional form
restrictions.

\subsubsection{Congestion charge}

Under a congestion charge, the targeted alternative(s) becomes less
attractive, but the equilibrium congestion may fall enough to improve the
payoff of those who continue to choose it. Demand diverted to untargeted
alternatives worsens their congestion. Let $\boldsymbol{\pi }$ and $%
\boldsymbol{\pi }^{\ast }$ denote the unique pre- and post-policy unique
equilibria, i.e.%
\begin{equation*}
\boldsymbol{\pi }_{j}^{\ast }={}\Pr \left[ \widetilde{U}_{j}(S_{ij},R_{j}^{%
\ast },\boldsymbol{\pi }^{\ast })>\max_{k\neq j}\widetilde{U}%
_{k}(S_{ij},R_{j}^{\ast },\boldsymbol{\pi }^{\ast })\right] \text{.}
\end{equation*}%
Compensating variation $T_{i}$ of the congestion charge is defined by 
\begin{equation}
\max_{j}\widetilde{U}_{j}(S_{ij},R_{j},\boldsymbol{\pi })=\max_{j}\widetilde{%
U}_{j}(S_{ij},R_{j}+T_{i},\boldsymbol{\pi }^{\ast }).  \label{eq:cv}
\end{equation}%
Because $\boldsymbol{\pi }^{\ast }$ is also unique by Theorem \ref%
{thm:unique}, $T_{i}$ is unambiguously defined.

\textbf{Comparative Statics}: We now show that increasing the price of one
alternative cannot increase its equilibrium share, but every other share may 
\textit{not} rise. Essentially, the first-round substitution toward
untargeted alternatives changes their congestion, producing second-round
substitution among them. This distinction is important in our application
because a charge at oversubscribed colleges changes offer probabilities
throughout the college system.

Toward that end, consider $J$ alternatives. Suppose $t_{j}=\left(
r_{j},p_{j}\right) $ and let $\pi =(\pi _{1},\ldots ,\pi _{J})$ denote an
equilibrium choice-probability vector under prices $p$, and let $\pi
^{\prime }=(\pi _{1}^{\prime },\ldots ,\pi _{J}^{\prime })$ denote an
equilibrium after the price of alternative 1 increases from $p_{1}$ to $%
p_{1}^{\prime }>p_{1}$, with all other prices unchanged.

\begin{proposition}
Assume that utility from alternative $j$, $U_{j}(r_{j},p_{j};\pi _{j})$ is
weakly decreasing in its own equilibrium probability $\pi _{j}$. Assume also
that an increase in $p_{1}$ weakly lowers the utility of alternative 1 for
every type, leaves the utilities of alternatives $j\neq 1$ unchanged, and
that ties occur with probability zero. If $p_{1}^{\prime }>p_{1}$, then the
equilibrium probability of alternative 1 cannot increase, i.e.%
\begin{equation*}
\pi _{1}^{\prime }\leq \pi _{1}.
\end{equation*}%
If, in addition, the increase in $p_{1}$ strictly lowers utility for a
positive mass of individuals who are marginal between alternative 1 and some
other alternative, then $\pi _{1}^{\prime }<\pi _{1}.$
\end{proposition}

\begin{proof}
Suppose, to the contrary, that $\pi _{1}^{\prime }>\pi _{1}.$Define $%
A=\{j:\pi _{j}^{\prime }>\pi _{j}\}$. Then $1\in A$ and $\sum_{j=1}^{J}\pi
_{j}=\sum_{j=1}^{J}\pi _{j}^{\prime }=1$. Hence 
\begin{equation*}
\sum_{j\in A}\pi _{j}^{\prime }>\sum_{j\in A}\pi _{j}.
\end{equation*}%
Therefore, a positive mass of individuals must switch from an alternative
outside $A$ before the price change to an alternative in $A$ after the price
change. Consider such an individual. Let $k\notin A$ be the alternative
chosen before the price change and let $j\in A$ be the alternative chosen
after the price change. Since $1\in A$, we have $k\neq 1$. Also,%
\begin{equation*}
\pi _{k}^{\prime }\leq \pi _{k}.
\end{equation*}%
Because the price of alternative $k$ is unchanged and utility is weakly
decreasing in own congestion, 
\begin{equation*}
U_{k}(r_{k},p_{k};\pi _{k}^{\prime })\geq U_{k}(r_{k},p_{k};\pi _{k}).
\end{equation*}%
Thus alternative $k$ is weakly more attractive after the change. For $j\in A$%
, we have $\pi _{j}^{\prime }>\pi _{j}$. If $j\neq 1$, its price is
unchanged and therefore 
\begin{equation*}
U_{j}(r_{j},p_{j};\pi _{j}^{\prime })\leq U_{j}(r_{j},p_{j};\pi _{j}).
\end{equation*}%
If $j=1$, both the increase in congestion and the increase in $p_{1}$ make
alternative 1 weakly less attractive, so 
\begin{equation*}
U_{1}(r_{1},p_{1}^{\prime };\pi _{1}^{\prime })\leq U_{1}(r_{1},p_{1};\pi
_{1}).
\end{equation*}%
Hence every alternative in $A$ is weakly less attractive after the price
change, while every alternative outside $A$ is weakly more attractive.
Because the individual chose $k$ before the change, 
\begin{equation*}
U_{k}^{\mathrm{old}}\geq U_{j}^{\mathrm{old}},
\end{equation*}%
while choosing $j$ after the change requires 
\begin{equation*}
U_{j}^{\mathrm{new}}\geq U_{k}^{\mathrm{new}}.
\end{equation*}%
But the monotonicity conditions imply 
\begin{equation*}
U_{j}^{\mathrm{new}}\leq U_{j}^{\mathrm{old}}\leq U_{k}^{\mathrm{old}}\leq
U_{k}^{\mathrm{new}}.
\end{equation*}%
Combining these inequalities with optimality after the change forces 
\begin{equation*}
U_{j}^{\mathrm{new}}=U_{k}^{\mathrm{new}},
\end{equation*}%
a tie. Since ties occur with probability zero, a positive mass of
individuals cannot switch from outside $A$ into $A$. This contradicts the
fact that the total equilibrium probability of the alternatives in $A$
increases. Therefore, 
\begin{equation*}
\pi _{1}^{\prime }\leq \pi _{1}.
\end{equation*}%
The strict version follows when the price increase strictly lowers the
utility of alternative 1 for a positive mass of individuals who are marginal
between alternative 1 and another alternative.
\end{proof}

\textbf{Numerical Illustration: }To show that the share of non-targeted
alternatives may rise or fall, consider the following numerical exercise
with three alternatives. Suppose the fixed point condition is $\pi =G(p,\pi
) $, with 
\begin{equation*}
D_{\pi }G=\left[ \frac{\partial G_{j}}{\partial \pi _{k}}\right]
_{j,k=1,2,3}=%
\begin{pmatrix}
-0.4 & 0.1 & 0.1 \\ 
0.3 & -0.2 & 0.1 \\ 
0.1 & 0.1 & -0.2%
\end{pmatrix}%
.
\end{equation*}%
The diagonal elements are negative and the off-diagonal elements are
positive. Each column sums to zero, consistent with 
\begin{equation*}
G_{1}+G_{2}+G_{3}=1.
\end{equation*}

Suppose the direct effect of an increase in $p_{1}$ is 
\begin{equation*}
G_{p_{1}}=%
\begin{pmatrix}
-0.10 \\ 
0.01 \\ 
0.09%
\end{pmatrix}%
.
\end{equation*}%
Thus, holding congestion fixed, a higher $p_{1}$ reduces the probability of
alternative 1 and increases the probabilities of alternatives 2 and 3. The
entries sum to zero. Differentiating the equilibrium conditions gives 
\begin{equation*}
\left( I-D_{\pi }G\right) \frac{d\pi }{dp_{1}}=G_{p_{1}}.
\end{equation*}%
Here 
\begin{equation*}
I-D_{\pi }G=%
\begin{pmatrix}
1.4 & -0.1 & -0.1 \\ 
-0.3 & 1.2 & -0.1 \\ 
-0.1 & -0.1 & 1.2%
\end{pmatrix}%
.
\end{equation*}%
Solving gives 
\begin{equation*}
\frac{d\pi }{dp_{1}}=%
\begin{pmatrix}
-0.06667 \\ 
-0.00256 \\ 
0.06923%
\end{pmatrix}%
.
\end{equation*}

Thus, $\frac{d\pi _{1}}{dp_{1}}<0$, in accordance with the proposition.
However, $\frac{d\pi _{2}}{dp_{1}}<0$ even though $\frac{\partial G_{2}}{%
\partial p_{1}}>0$. Hence the direct substitution effect toward alternative
2 is more than offset by equilibrium congestion feedback.

\section{Cambridge College Applications}

Cambridge and Oxford are two of the world's oldest and most prestigious
universities, and undergraduate admissions to them are intensely competitive
(cf. Bhattacharya, Kanaya and Stevens 2014, Bhattacharya and Shvets 2024). A
candidate is not allowed to apply to both in the same year. Here, we focus
on an aspect of the Cambridge applications process which provides an
empirical counterpart to our theoretical model. An undergraduate applicant
in a given subject $s$ e.g. history, economics, math etc., chooses and
applies through one Cambridge college which makes the final admission
decision on him/her. An applicant cannot apply to more than one Cambridge
college in the same year.\footnote{%
A small number of students make an `open' application, in which case they
are randomly allocated to a college.} The colleges differ in location,
accommodation cost, size, age, endowment, and the number of fellows in
particular subjects. Capacity within each college in each subject of study
is determined centrally \textit{before} applicants make their choices.
Applicants can access publicly available information on colleges and
historical admission statistics (see last table below for an example). This
enables them to choose based on anticipated congestion which, together with
their academic credentials $Z_{i}$, determines the eventual probability of
an offer at college $j$, denoted by $O_{sj}$. At the initial application
stage, an additional applicant to college $j$ competes most directly for
places at $j$. In a number of cases, an applicant who is unsuccessful at her
chosen college can be admitted by another college through the Cambridge
pool. Such a channel is permitted by Assumption \ref{ass:negative}, as shown
in the Appendix below. In our empirical work, we take the initial
application college as the college of choice. This is because in the
official information available to applicants (see the last table
below),\textquotedblleft ...the college of application in the data is
defined as the applicant's first choice college (or, in the case of an open
applicant, the college to which the applicant was allocated) regardless of
subsequent transfers through the pools.\textquotedblright\ There is no
public information available exclusively on the pool. So it is practically
impossible for applicants to guess if applying to a specific college can
improve their chances of eventual acceptance \textit{via the pool} beyond
the aggregate success probabilities of the type reported in the very last
table below. For this reason, we do not include separate considerations of
the pool into our specification of individual choice below.

The variation in college characteristics by subject is useful for
identification. Indeed, capacity, distance to the relevant department, the
number of fellows etc. vary by subject within a college, whereas many
unobserved amenities of the college (e.g. how nice the college looks) are
common across subjects. The data therefore contain repeated choice
environments with a common college component but different congestion and
observed subject-specific attributes.

Our analysis focuses on applicant sorting, taking college capacities and
admissions criteria as given while allowing admission cutoffs to adjust to
application demand. Given fixed capacity, these adjustments determine the
offer rates $O_{sj}$, and the associated probabilities $\pi _{sj}$, in
equilibrium. Individual applicants take these probabilities as given when
choosing a college, just as individual consumers in empirical demand
settings take prices as given, and prices are determined in equilibrium by
aggregate market behavior. Historical acceptance rates, publicly available
on Cambridge's website and reasonably stable over time, provide information
on admission prospects (see the official admissions statistics reproduced at
the end of the tables and figures). Thus, we capture the college side
through the mapping from application patterns to admission probabilities,
rather than through a separate model of colleges' objectives or capacity
decisions. Our welfare estimates thus have a short-run interpretation: they
incorporate equilibrium adjustments in applicant sorting and admission
probabilities \emph{while holding capacities fixed}. This approach follows
the tradition of welfare analysis based on estimated demand without a full
model of supply (cf. Lewbel, 2001; Hausman and Newey, 2016, 2017;
Bhattacharya, 2015, 2018). Also, in what follows, we sometimes drop the $s$
subscript to avoid notational clutter when there is no chance of confusion.

\section{Empirical Model and Identification}

An applicant indexed by $i=1,...I$, chooses exactly one
out of $J$ colleges, indexed by $j=1,...J$. The pay-off of applicant $i$ 
\textit{upon having chosen college }$j$ equals%
\begin{equation*}
\left\{ 
\begin{array}{l}
H_{ij}\text{, if admitted,} \\ 
0\text{, if rejected.}%
\end{array}%
\right.
\end{equation*}%
We make the following assumptions.

\begin{assumption}
$H_{ij}>0$ with probability 1, i.e., upon having chosen college $j$, the
pay-off from rejection is lower than the payoff from acceptance for each $%
j=1,...,J$.
\end{assumption}

\begin{remark}
Note that an applicant $i$ can also apply to and get admitted in
universities other than Cambridge. The above assumption allows her payoff $%
K_{i}$ from that outcome to satisfy that for two different Cambridge
colleges $j\neq k$, $H_{ij}<K_{i}<H_{ik}$, and/or%
\begin{equation*}
p_{j}H_{ij}<K_{i}<p_{k}H_{ik}\text{,}
\end{equation*}%
where $p_{j}$ is the probability of admission. While a student can apply to
a single Cambridge college, there is no restriction on whether they
simultaneously apply to universities other than Cambridge.\footnote{%
A student cannot apply to Oxford and Cambridge in the same year.}
\end{remark}

\begin{assumption}
Applicant $i$'s perceived probability of a successful outcome upon choosing
college $j$ equals $\tilde{O}_{j}\left( \pi _{ij}^{e},Z_{i}\right) $ taking
values in $\left[ 0,1\right] $, where $\pi _{ij}^{e}$ is the equilibrium
proportion of candidates choosing college $j$ as perceived by $i$, $Z_{i}$
denotes applicant specific characteristics observed by the college, by the
researcher and the applicant, $\tilde{O}_{j}\left( \pi
_{ij}^{e},Z_{i}\right) $ is weakly decreasing in $\pi _{ij}^{e}$. When
applying for admissions, an applicant $i$ knows his/her own $Z_{i}$ but not
the $Z$ of other applicants.
\end{assumption}

Assumption 4 clarifies the source of negative externality in our model.
Firstly, the perceived offer probability is calculated as a function of own
characteristics and the fraction choosing option $j$. As $\pi _{ij}^{e}$
rises, the perceived probability of acceptance for applicant $i$ with any
specific value of $Z_{i}$ goes down. This happens due to capacity constraint
that restricts the number of places in each college. Further, more people
applying to college $k$ has no effect on the perceived expected utility of a
candidate applying to college $j\neq k$, i.e. we are in the special case
outlined in Remark 1 above. This is because (A) in the information available
to students (see extract from official statistics at the end of the last
table below), the college of application is defined as the applicant's first
choice college regardless of subsequent transfers through the pools, and (B)
no separate information exists on the differential rates of admission
through the pool across first-choice colleges.

\begin{assumption}
The applicants perceive the $\pi $'s correctly. That is, $\pi _{ij}^{e}=\pi
_{j}$, the true equilibrium proportion of applicants choosing college $j$,
and $\tilde{O}_{j}\left( \pi _{ij}^{e},Z_{i}\right) =\tilde{O}_{j}\left( \pi
_{j},Z_{i}\right) $, with $\tilde{O}_{j}\left( \cdot ,Z_{i}\right) $ weakly
decreasing for any fixed value of $Z_{i}$.
\end{assumption}

\begin{assumption}
Let $n=\sum_{j=1}^{J}n_{j}$, the total capacity available, aggregated across
all colleges, and assume, as is the case empirically, $n<<I$.
\end{assumption}

Thus the expected utility of applicant $i$ from choosing college $j$ is
given by $\tilde{O}_{j}\left( \pi _{j},Z_{i}\right) \times H_{ij}$ where $%
H_{ij}>0$ with probability 1 by Assumption 3. Then individual $i$ prefers
alternative $j$ over $k$ if and only if%
\begin{equation*}
\tilde{O}_{j}\left( \pi _{j},Z_{i}\right) \times H_{ij}>\tilde{O}_{k}\left(
\pi _{k},Z_{i}\right) \times H_{ik}\text{.}
\end{equation*}%
So, the aggregate probability that option $j$ is chosen is given by%
\begin{equation}
\pi _{j}=\Pr \left[ \tilde{O}_{j}\left( \pi _{j},Z_{i}\right) \times
H_{ij}>\max_{k\neq j}\left\{ \tilde{O}_{k}\left( \pi _{k},Z_{i}\right)
\times H_{ik}\right\} \right] \text{,}  \label{0'}
\end{equation}%
under the assumption that probability of an exact tie is zero. Since $\tilde{%
O}_{j}\left( \pi _{j},Z_{i}\right) $ is weakly decreasing in $\pi _{j}$, and 
$U_{ij}>0$ with probability 1, this is precisely of the form (\ref{eq:G})
with $\tilde{U}_{ij}=\tilde{O}_{j}\left( \pi _{j},Z_{i}\right) \times H_{ij}$
and therefore by Theorem 1, entails a unique equilibrium $\mathbf{\pi }$.

For ease of econometric modelling, we further assume that the applicant's 
\textit{own perceived} probability of receiving an offer takes a
multiplicative form 
\begin{equation}
\widetilde{O}_{j}(\pi _{j};Z_{i})=O_{j}(\pi _{j})\phi _{j}(Z_{i}),
\label{eq:offer-separable}
\end{equation}%
where $O_{j}(\cdot )$ is weakly decreasing. Here $\phi _{j}(Z_{i})$ is a
baseline subjective measure of academic ability based on the applicant's
past credentials., satisfying $0<\phi _{j}(Z_{i})<1$. The term $O_{j}(\pi
_{j})=\frac{n_{j}}{\max \left\{ n_{j},I\pi _{j}\right\} }$, where $n_{j}$ is
capacity and $I$ is the total number of applicant, measures how likely she
thinks this ability would secure her admission, depending on how many others
have applied to the college. If $n_{j}>I\pi _{j}$, then the offer rate is 1.
The colleges can leave spots vacant if there are not sufficiently many
qualified candidates. In our empirical exercise, $n_{j}<I\pi _{j}$ in every
case.

\begin{remark}
Note that applicant $i$ does not observe the $Z$ for other applicants either
in the current or in past years; so she cannot predict her \textit{relative}
ranking, nor can she predict how students of different abilities will sort
differentially following an intervention.
\end{remark}

We now re-introduce the $s$ subscript to denote subject. Let $W_{sj}$ denote
observed characteristics that vary across both subjects and colleges, $R_{j}$
observed college characteristics that do not vary across subjects, and $%
Z_{i} $ observed applicant characteristics. Let $\xi _{j}$ denote a college
feature observed by applicants but not by the econometrician. Examples
include how nice the college looks and how friendly it appears during the
open day. Finally, $\varepsilon _{sij}$ is idiosyncratic preference
heterogeneity. In what follows, we sometimes drop the $s$ subscript to avoid
notational clutter when there is no chance of confusion. We impose a linear
index structure on \textit{log} utility: 
\begin{equation}
\ln \phi _{js}(Z_{i})+\ln H_{j}(W_{sj},R_{j},Z_{i},\xi _{j},\varepsilon
_{sij})=W_{sj}^{\prime }\gamma +R_{j}^{\prime }\alpha +Z_{i}^{\prime }\delta
_{sj}+\xi _{j}+\varepsilon _{sij}.  \label{eq:indexprimitive}
\end{equation}%
This specification preserves positivity of the underlying payoff.

Define%
\begin{equation}
\bar{\xi}_{j}=R_{j}^{\prime }\alpha +\xi _{j}.  \label{eq:xibar}
\end{equation}%
Then the choice probability for college $j$ is generated by comparison of 
\begin{equation}
\ln O_{j}(\pi _{sj})+W_{sj}^{\prime }\gamma +Z_{i}^{\prime }\delta _{sj}+%
\bar{\xi}_{j}+\varepsilon _{sij}  \label{eq:index}
\end{equation}%
across colleges. Taking college $J$ as the reference, write the vector of
systematic utility contrasts as 
\begin{equation}
t_{sij}=A_{sj}+Z_{i}^{\prime }B_{sj},
\end{equation}%
where 
\begin{equation}
A_{sj}=(W_{sj}-W_{sJ})^{\prime }\beta +(\gamma _{j}-\gamma _{J})+\ln
O_{j}(\pi _{sj})-\ln O_{J}(\pi _{sJ}),
\end{equation}%
\begin{equation}
B_{sj}=\delta _{sj}-\delta _{sJ},\qquad \gamma _{j}=R_{j}^{\prime }\alpha
+\xi _{j}.
\end{equation}

Identification proceeds in two stages. The first concerns recovery of these
indices from conditional choice probabilities. The second concerns
decomposition of the identified indices into observed characteristics,
congestion, and college-specific unobservables.

\textbf{First-stage identification.} With a known choice-probability link
function, identification of the indices follow from inversion and
appropriate regressor rank conditions. With an unknown link, however,
invertibility alone is insufficient: the data must also distinguish changes
in index coefficients from changes in the link function. Semiparametric
approaches such as Ahn et al.\ 2018 or Khan et al 2021 therefore require
specific support and rank conditions. Accordingly, our decomposition
argument is conditional on the following first-stage requirement.

\begin{condition}
Conditional choice probabilities identify $A_{sj}$, $B_{sj}$ for each
subject $s$ and for $j=1,...,J$ under location and scale restrictions that
make the indices comparable across subjects.
\end{condition}

The above condition separates the requirements for identifying the
underlying choice model from the additional restrictions used below. It can
be guaranteed by a correctly specified parametric choice model with
appropriate rank conditions, or by a semiparametric identification result
whose necessary support and rank conditions are satisfied by the available
variation in observables, as outlined below.

\textbf{A route to semiparametric identification:} The first-stage condition
above can, in principle, be established without specifying a parametric
distribution for preference heterogeneity. In the invertible-index framework
of Ahn et al.\ 2018 or Khan et al 2021, conditional choice probabilities
identify the normalized index coefficients under appropriate support and
conditional-variation conditions. Once these coefficients are identified,
the choice-probability link is also identified on the support of the index
vector. In particular, consider our additive log-utility specification.
Suppress the subject subscript and write utility relative to reference
option $J$ as 
\begin{equation*}
t_{ij}+\eta _{ij},\qquad \eta _{ij}=\varepsilon _{ij}-\varepsilon
_{iJ},\qquad j\neq J.
\end{equation*}%
If the vector $\eta _{i}$ is independent of the regressors, the probability
of choosing the reference college satisfies 
\begin{equation*}
P_{J}(t)=\Pr \left( \eta _{ij}\leq -t_{j}\mbox{ for all }j\neq J\right)
=F_{\eta }(-t).
\end{equation*}%
Thus, identification of the choice-probability link identifies the joint
distribution function of shock differences at the corresponding points in
its domain. Full support of the normalized index vector on the entire $(J-1)$%
-dimensional Euclidean space is sufficient to recover the entire
distribution.

These are distinct requirements: the conditional-variation conditions
identify the index coefficients, while the extent of index support
determines how much of the heterogeneity distribution -- and hence welfare
distribution -- can be recovered.

\textbf{Decomposition across subjects.} Suppose the first-stage condition
above holds. For a fixed college $j$, subtracting the identified intercept
for subject $r$ from that for subject $s$ eliminates the common college
component: 
\begin{eqnarray}
A_{sj}-A_{rj} &=&\left\{ (W_{sj}-W_{sJ})-(W_{rj}-W_{rJ})\right\} ^{\prime
}\beta  \notag \\
&&+\left\{ \ln O_{j}(\pi _{sj})-\ln O_{J}(\pi _{sJ})\right\}  \notag \\
&&-\left\{ \ln O_{j}(\pi _{rj})-\ln O_{J}(\pi _{rJ})\right\} .
\end{eqnarray}%
Thus, full column rank of the stacked cross-subject differences in observed
characteristics identifies $\beta $. Substituting into the intercept
equation then identifies 
\begin{equation}
\gamma _{j}-\gamma _{J}\equiv (R_{j}-R_{J})^{\prime }\alpha +(\xi _{j}-\xi
_{J}).  \label{A}
\end{equation}%
Separating $\alpha $ from the college unobservables requires an additional
restriction. We impose either (A) or (B) below.

\textbf{(A) Orthogonality:} The observed college characteristics $R_{j}$ are
orthogonal to the unobserved college component $\xi _{j}$, conditional on $%
W_{sj}$'s. The matrix of relative college characteristics $R_{j}-R_{J}$ has
full column rank, with the corresponding conditional moment matrix
nonsingular.

\textbf{(B) Correlated random effects:} The unobserved college component
satisfies 
\begin{equation}
\xi _{j}=\overline{W}_{j}^{\prime }\varphi +\nu _{j},\qquad %
\mathop{\mathrm{Cov}}(\nu _{j},R_{j}\mid \overline{W}_{j})=0,\qquad 
\overline{W}_{j}=\frac{1}{S}\sum_{s=1}^{S}W_{sj}.
\end{equation}%
We additionally require the corresponding nonsingularity condition for
identifying $\alpha $ after controlling for $\overline{W}_{j}$. In
particular, the college characteristics must retain sufficient variation
after those controls are removed.

Under (A), orthogonality separates the contribution of observed college
characteristics from the unobserved college component. Under (B), the
correlation is accounted for through the cross-subject means $\overline{W}%
_{j}$, and the remaining variation in $R_{j}$ identifies $\alpha $. Once $%
\alpha $ is recovered, equation (24) identifies $\xi _{j}-\xi _{J}$. We
summarize the above discussion as a proposition:

\begin{proposition}
\textbf{\ (Identification conditional on first-stage index recovery).} 
\textit{Suppose the first-stage identification condition holds, the
conditional offer probabilities are observed , and the college component $%
\gamma _{j}$ is common across subjects. If the stacked cross-subject
differences in $W_{sj}-W_{sJ}$ have full column rank, then $\beta $, the
relative applicant-characteristic coefficients $B_{sj}$, and the relative
college components $\gamma _{j}-\gamma _{J}$ are identified. If, in
addition, restriction (A) or (B) holds with its associated rank condition,
then $\alpha $ and the relative college unobservables $\xi _{j}-\xi _{J}$
are identified under the maintained normalizations.}\label{prop:id}
\end{proposition}

Applicant variation identifies the within-subject slopes; repeated subjects
remove college fixed effects and identify $\gamma $; and an explicit
orthogonality or correlated-random-effects restriction separates observed
time-invariant college characteristics from unobserved ones. The
repeated-subject structure is especially useful here because unobserved
features such as the atmosphere or appearance of a college are plausibly
common across subjects even when congestion and subject-specific academic
characteristics vary.

This use of repeated subjects differs from the conventional strategy of
requiring an excluded demand shifter for every endogenous alternative
characteristic. The $\xi _{j}$ is a realized college-specific unobservable.
Because the same realized $\xi _{j}$ enters several subject choice problems,
differencing across subjects removes it before the final decomposition in (%
\ref{A}). The approach therefore exploits a panel-like structure across
choice environments. What must remain stable is the college component
observed by applicants but omitted from the econometrician's data; what must
vary is the subject-specific component that identifies $\gamma $.

This distinction is also important for welfare analysis. Recent work (cf.
Bhattacharya et al 2024) on discrete choice with social interactions shows
that welfare can fail to be identified when choice data reveal only
differences in spillover effects but welfare depends on their levels. Our
centralized-allocation counterfactual avoids that problem because welfare
loss can be expressed entirely in relative utilities. The congestion-charge
formula also has the same property once a non-targeted college is used as
the reference.

\subsection{Which welfare effects are identified?}

Identification of utility differences does not automatically imply
identification of every welfare counterfactual. The relevant question is
whether the intervention changes utility through objects whose levels and
counterfactual values can be recovered from the data.

\textbf{Centralized Allocation}: Suppress the subject subscript. Let $\pi
^{D}$ denote the decentralized equilibrium application shares, and let $%
p_{j} $ denote the probability of assignment to college $j$ under
centralized allocation. Write $O_{j}^{D}=O_{j}(\pi _{j}^{D})$ and $%
O_{j}^{C}=O_{j}(p_{j}) $ for the corresponding offer probabilities. We
assume these probabilities are positive on the alternatives considered.
Preferences and the distribution of applicant characteristics remain
unchanged across regimes.

Let $R_{j}$ denote rent and $\alpha <0$ its coefficient in log utility.
Define the baseline utility contrast 
\begin{equation}
a_{ij}^{D}=\frac{O_{j}^{D}\exp \!\left( W_{j}^{\prime }\beta +\alpha
R_{j}+Z_{i}^{\prime }\delta _{j}+\xi _{j}\right) }{O_{J}^{D}\exp \!\left(
W_{J}^{\prime }\beta +\alpha R_{J}+Z_{i}^{\prime }\delta _{J}+\xi
_{J}\right) },\qquad a_{iJ}^{D}=1.  \label{eq:baseline-relative-utility}
\end{equation}%
Also write $\eta _{ij}=\varepsilon _{ij}-\varepsilon _{iJ}$, with $\eta
_{iJ}=0$, and $\rho _{j}=O_{j}^{C}/O_{j}^{D}$. The above discussion on
identification of preferences implies that the utility contrasts $a_{ij}^{D}$
are point identified.

Compensating variation $T_{i}$ is the common reduction in rent under
centralized allocation that makes applicant $i$ indifferent between the two
regimes. Thus $T_{i}>0$ denotes a welfare loss from centralization. Because
reducing rent by $T_{i}$ multiplies each admission payoff by $\exp (-\alpha
T_{i})$, the indifference condition is 
\begin{equation*}
\max_{j}\{O_{j}^{D}H_{ij}\}=\exp (-\alpha T_{i})\sum_{j}p_{j}O_{j}^{C}H_{ij}.
\end{equation*}%
Here $H_{ij}$ is the admission payoff evaluated at the original rent.
Dividing both sides by the baseline expected payoff at the reference college
and taking logarithms gives 
\begin{equation}
T_{i}=\frac{\displaystyle\ln \max_{j}\!\left\{ a_{ij}^{D}e^{\eta
_{ij}}\right\} -\displaystyle\ln \!\left[ \sum_{j}p_{j}\rho
_{j}a_{ij}^{D}e^{\eta _{ij}}\right] }{-\alpha }.
\label{eq:centralization-cv}
\end{equation}

Equation~\eqref{eq:centralization-cv} separates the loss of individual
choice from the change in admission probabilities. The first term evaluates
optimal choice at the decentralized equilibrium; the second evaluates the
allocation lottery at the offer probabilities induced by centralization. The
common, unidentified level of utility cancels. Consequently, the
distribution of $T_{i}$ is identified when $a_{ij}^{D}$ are identified, and
the joint distribution of $\varepsilon _{ij}-\varepsilon _{iJ}$ are
identified or specified.

\textbf{Congestion Charge}: For a congestion charge, the counterfactual
index changes through the known price-change and through the resulting
equilibrium offer probability $O_{j}(\pi _{sj}^{\ast })$. Importantly, the
welfare effect arises not only from the direct charge. An applicant who
remains at a targeted college pays the charge but benefits from the
endogenous improvement in its offer probability; an applicant who switches
avoids the charge but contributes to and suffers from congestion elsewhere.
Now, define $a_{ij}^{\ast }$ analogously using the counterfactual rent
vector and the unique equilibrium shares $\pi _{j}^{\ast }$. Taking a
reference college $J$ that is not targeted, compensating variation can be
written as 
\begin{equation}
T_{i}=\frac{\max_{j}\{\ln a_{ij}^{D}+\varepsilon _{ij}-\varepsilon
_{iJ}\}-\max_{j}\{\ln a_{ij}^{\ast }+\varepsilon _{ij}-\varepsilon
_{iJ}\}+\ln \pi _{J}^{\ast }-\ln \pi _{J}}{-\alpha }.  \label{eq:chargeCV}
\end{equation}%
where $\alpha $ is the coefficient on price. Thus the compensating variation
depends only on identified utility contrasts, the shock differences, and the
identified counterfactual shares, and is therefore identified once the joint
distribution of $\varepsilon _{ij}-\varepsilon _{iJ}$ are identified or
specified.

\begin{remark}
\textit{Note that the exponential terms in (\ref{eq:centralization-cv}) and (%
\ref{eq:chargeCV}) expression come from the multiplicative expected-utility
specification, not from an extreme-value assumption.}
\end{remark}

\textbf{Change in Individual Covariate}: A counterfactual change in an
individual covariate $Z_{i}$, by contrast, generally changes the level of
utility at every college. Choice probabilities identify $B_{sj}=\delta
_{sj}-\delta _{sJ}$, not all of the levels $\delta _{sj}$. Adding the same
coefficient vector to every $\delta _{sj}$ does not change any observed
choice probability but does change the welfare effect of changing $Z_{i}$.
Consequently, without an additional restriction fixing this common
component, the distribution of compensating variation from a change in $%
Z_{i} $ is \textit{not} point identified. This provides a useful contrast:
equilibrium uniqueness makes a counterfactual allocation single valued, but
welfare is point identified only when the utility changes induced by the
policy are themselves identified.

Note that the welfare expressions above depend on relative systematic
utilities, shock differences, the numeraire coefficient, and counterfactual
offer probabilities. They therefore admit \textit{semiparametric}
identification when the index-identification conditions hold, the available
support identifies the joint distribution of shock differences needed for
the welfare calculation, and the numeraire coefficient and counterfactual
offer probabilities are identified under the maintained structural
restrictions. The cancellation of a common utility level removes one
potential obstacle to identification; it does not remove these support and
distributional requirements.

\section{Data and Empirical Results}

We use administrative micro-data for UK-based Cambridge applicants from
2013--2017 for six large subjects, containing 25,325 applicants: 16,394 men
and 8,931 women. These subjects jointly account for a large share of
Cambridge undergraduates and include Economics, Engineering, Mathematics,
Law, Medicine, and Natural Sciences. We merge applicant records with college
characteristics assembled from college and university sources. College
attributes include size of the undergraduate body, endowment, median weekly
rent, age, mean academic ranking in Cambridge examinations, distance from
the college to the applicant's proposed department, the number of fellows in
the applicant's subject, and the number of offers historically made to
applicants from the same school type. Applicant covariates include sex,
prior academic achievements and state/private school status. Summary
statistics appear in Table 1.

Offer rates and numbers of places vary by college, subject, and year.
Applicants can access past statistics on colleges via publicly available
admission statistics (see extract at the end of the tables and figures
below).

We use the data pooled across years to estimate preference parameters, while
welfare counterfactuals use the relevant year- and subject-specific
capacities and offer rates. Pooling supplies the cross-subject variation
needed for preference identification, whereas the counterfactual equilibrium
respects the actual congestion due to fixed capacity in the college-subject
and year being analyzed.

In what follows, we maintain conditional independence, i.e. Assumption (A)
above. Further, for the empirical implementation, we assume independent,
identically distributed type-I extreme-value utility shocks, independent of
the observed covariates, with the maintained location and scale
normalizations. This yields multinomial logit choice probabilities (McFadden
1973): 
\begin{equation}
P_{sij}=\frac{\exp \{\ln O_{j}(\pi _{sj})+W_{sj}^{\prime }\gamma
+R_{j}^{\prime }\alpha +Z_{i}^{\prime }\delta _{j}+\xi _{j}\}}{%
\sum_{k=1}^{J}\exp \{\ln O_{k}(\pi _{sk})+W_{sk}^{\prime }\gamma
+R_{k}^{\prime }\alpha +Z_{i}^{\prime }\delta _{k}+\xi _{k}\}}.
\label{eq:logit}
\end{equation}%
This restriction serves both an identification and a practical computational
role. The available applicant characteristics include discrete measures of
academic attainment and school type. Their variation, together with
variation across subjects and years, may not satisfy the support conditions
needed to recover an unrestricted joint distribution of preference
heterogeneity over its entire support. The logit specification supplies a
distributional structure that permits out-of-sample computation of the
counterfactual equilibria and welfare distributions. Secondly, computing
counterfactual equilibria requires many iterations of coefficient
estimation; the logit simplifies this exercise. We emphasize that the
equilibrium uniqueness theorem and the analytical welfare comparison
expressions do \textit{not} depend on this distributional assumption. In
settings with large support of explanatory variables, the semiparametric
approach, as in Ahn et al 2018 or Khan et al 2021, can be employed instead.
McFadden's pseudo $R^{2}$ ranges between $0.17$ and $0.32$ corresponding to
the estimates we obtain below in Table 2, suggesting a decent goodness of
fit.

Also, one can impose the equilibrium restrictions 
\begin{equation}
\pi _{sj}=\int P_{sij}(\boldsymbol{\pi }_{s})\,dF_{s}(Z),\qquad j=1,\ldots
,J,  \label{eq:estfixedpoint}
\end{equation}%
jointly with the likelihood. Such an estimator may be more efficient.
However, in the empirical exercise, we instead use the observed choice
shares in the likelihood and perform the fixed-point solution for the
congestion-charge counterfactual. This avoids adding a computational layer
to estimation while preserving the equilibrium feedback where it is
essential for policy analysis.

Table 2 reports the coefficients on observed college characteristics under
the multinomial logit. We estimate the model separately by sex, use open
applications (where students do not specify a college) as the base
alternative, and obtain standard errors by bootstrap. For the
applicant-characteristic slopes, we allow the effects of GCSE performance
and school type on college choice to differ across colleges. The
college-specific coefficients on individual characteristics (the $\delta
_{j} $s above), pooled across subjects and years, are reported in Table 3A-B
below.

It is apparent that applicants of both sexes prefer colleges closer to their
academic department, with more fellows in their subject, and with a larger
historical number of offers to applicants of their school type. Rent enters
negatively for both sexes. The specification also provides direct evidence
that applicants respond to congestion: a test of a zero coefficient on the
log offer rate against the expected-utility normalization of one rejects the
zero-coefficient specification at the five-percent level for both men and
women. These estimates provide the preference inputs for the counterfactual
equilibria.

\subsection{Centralized random allocation}

For a switch from decentralized college choice to centralized random
allocation, we express welfare changes as compensating variation in weekly
rent. A positive number is a welfare loss: it is the rent reduction required
to compensate an applicant for the intervention. Figure 1 plots the
distribution of compensating variation for men and women. Losses are
positive for both groups, and the distribution of losses for men first-order
stochastic dominates that for females. The positive support is explained by
the fact that the ability to choose is eliminated but congestion is not
reduced because a rise in probability of being allocated to a college is
exactly offset by a proportional decrease in being admitted there since more
applicants will be allocated there in expectation. Note that this is not
specific to \textit{proportional} allocation. Any centralized allocation
scheme (i.e. where allocation to option $j$ is made with probability $%
p_{j}\neq \frac{n_{j}}{n}$) would lead to exactly the same cancellation i.e.
for $\pi _{j},p_{j}\in \left[ 0,1\right] $,%
\begin{equation*}
\underset{\text{Proportional allocation}}{\underbrace{\frac{n_{j}}{n}\times 
\frac{n_{j}}{I\frac{n_{j}}{n}}}}=\frac{n_{j}}{I}=\underset{\text{%
Decentralized}}{\underbrace{\pi _{j}\times \frac{n_{j}}{I\pi _{j}}}}=%
\underset{\text{Any centralized allocation}}{\underbrace{p_{j}\times \frac{%
n_{j}}{Ip_{j}}}}
\end{equation*}

Table 4 shows that the median loss is \pounds 94.84 for men and \pounds %
61.54 for women, and these differences are statistically significant.

This happens for two distinct reasons. Firstly, males have more
heterogeneous utilities and hence higher welfare loss from the inability to
choose (e.g. for $\lambda \in \left[ 0,1\right] $, $\max \left\{ a,b\right\}
-\left( \lambda a+\left( 1-\lambda \right) b\right) $ is larger when $a$ and 
$b$ are farther). Secondly, the coefficient on rent for males is smaller in
absolute value (see Table 2 below), hence a larger compensation is required
for the same loss in utility.

\subsection{Congestion charge}

We next consider a counterfactual increase in weekly rent by \pounds 100 at
King's and Trinity, two highly popular colleges. The amount is chosen to be
deliberately large to take our calculations `out-of-sample'. The
intervention directly lowers the utility of applying to the targeted
colleges but reduces their application shares and therefore improves offer
probabilities for applicants who continue to choose them. Demand diverted to
other colleges would potentially raise congestion elsewhere reducing
utility. For Economics in 2013, the full set of equilibrium shares before
and after the congestion charge appears in Table 6 below, which shows that
female application to Trinity and King's drops more than male and the
reduction is spread differentially across non-targeted colleges. For
example, at Gonville and Caius, male application rates go down whereas
female applications go up as a result of congestion charges at Trinity and
King's. This illustrates the theoretical point made above, regarding the
ambiguity of comparative statics predictions related to a congestion charge.

Figure 2 below plots the distribution of welfare loss by gender. Figures 3A
and 3B compare the two interventions for men and women, respectively. It is
clear that the welfare effects have supports that include both positive and
negative values, indicating positive probabilities of both losses and gains,
respectively. The welfare-loss quantiles are reported in Table 5.

We see substantial heterogeneity in overall welfare effects. For men, the
median compensating variation (we do not calculate the mean due to its
well-known sensitivity to outliers) is a loss of $\pounds 98.27$. For women,
the median is $-\pounds 27.59$, a welfare \textit{gain}. Moreover, nearly $%
75\%$ of men suffer welfare losses, while no more than $25\%$ of women do.
Given that all coefficients (cf. Tables 2, 3A-B) and changes in
congestion-charge led changes in choice probabilities (see Table 6) differ
by sex, it is not possible to pinpoint (cf. eqn. (\ref{eq:chargeCV})) a
single reason for the sex-difference. Aside from the gender-difference in
the rent coefficients alluded to above, a second possible explanation is
suggested by Table 6, which shows (cf. the highlighted entries) that a
larger fraction of women would leave those targeted colleges whose
characteristics matters relatively less to them.

Note also that the contrast between the two \textit{interventions} is
informative. Centralized allocation removes choice and, by design, does not
remove congestion effects ex ante, thus producing welfare loss for both
groups. A targeted congestion charge enables choice while altering the
congestion-related incentives. It can therefore improve welfare for a
subgroup whose demand is sufficiently concentrated at the targeted colleges,
even though it creates losses for others. The result illustrates (A) why the
equilibrium response, rather than the direct price effect alone, is
essential for welfare analysis with congestion, and (B) that even with a
negative externality, reducing demand for the most congested alternatives
need not be Pareto improving, and its net welfare effect can differ in sign
across subgroups.

\section{Incorporating Behavioral Frictions}

Our baseline framework assumes rational expectations, implying that
decision-makers accurately perceive the objective, equilibrium success
probabilities associated with each choice alternative. However, a
cornerstone of the modern behavioral public economics literature emphasizes
that individuals operating in complex, high-stakes environments frequently
suffer from systemic informational frictions and cognitive biases (Chetty et
al., 2009; Stantcheva, 2021). In capacity-constrained sorting markets,
applicants may systematically misperceive their true admission or access
probabilities due to bounded rationality or inadequate access to historical
institutional data.

To evaluate the efficiency and equity consequences of policy interventions
in these environments, an empirical model must be capable of disentangling
structural preference parameters from subjective behavioral distortions.
Following the classic probability-weighting literature (Prelec, 1998), we
now relax the rational expectations assumption by assuming that
decision-makers do not optimize with respect to the objective equilibrium
choice share vector $\pi $. Instead, they map true probabilities into
subjective, decision-relevant probabilities via a continuous monotonic
transformation mechanism.

Specifically, we assume that the objective success probability $O_{sj}$ for
alternative $j$ in market $s$ is transformed into a subjective,
decision-relevant probability $\tilde{O}_{sj}(O_{sj};\gamma )$ governed by
the structural parameter $\gamma >0$: 
\begin{equation}
\tilde{O}_{sj}(O_{sj};\gamma )=\exp \left( -(-\ln O_{sj})^{\gamma }\right)
\end{equation}%
The benchmark case of $\gamma =1$ characterizes standard rational
expectations. Values of $\gamma <1$ yield the classic inverse-S-shaped
probability weighting function, capturing a behavioral tendency to
over-weight small probabilities and under-weight high probabilities.

Under this behavioral specification, the log systematic decision utility
derived by individual $i$ from choosing alternative $j$ in market $s$ is
formulated as a linear index: 
\begin{equation}
U_{isj}^{D} = g(O_{sj}; \gamma) + W_{sj}^{\prime }\beta + Z_{i}^{\prime
}\delta_{sj} + R_{j}^{\prime }\alpha + \xi_{j} + \epsilon_{isj}
\end{equation}
where $g(O_{sj}; \gamma) \equiv \ln \tilde{O}_{sj}(O_{sj}; \gamma) = -(-\ln
O_{sj})^{\gamma}$, and the remaining components denote observed
market-varying characteristics ($W_{sj}$), individual attributes ($Z_{i}$),
invariant option features ($R_{j}$), unobserved public amenities ($\xi_{j}$%
), and standard idiosyncratic preference shocks ($\epsilon_{isj}$).

\textbf{Decision Utility versus Experienced Welfare: }An important
implication of the behavioral integration is that it formalizes the
conceptual wedge between \textit{decision utility}---which drives observed
choices in the data---and \textit{experienced welfare utility}---which
dictates the actual expected utility realized ex-post under objective match
probabilities. Let $j_{i}^{B}$ denote the optimal choice targeted by
individual $i$ under distorted subjective beliefs: 
\begin{equation}
j_{i}^{B}=\arg \max_{j\in \{1,\dots ,J\}}\left\{ g(O_{sj};\gamma
)+W_{sj}^{\prime }\beta +Z_{i}^{\prime }\delta _{sj}+R_{j}^{\prime }\alpha
+\xi _{j}+\epsilon _{isj}\right\}
\end{equation}%
Because the true probability of success is dictated by the physical,
equilibrium capacity constraints of the market rather than the agent's
internal perceptions, the experienced ex-ante welfare utility realized by
the agent under the status quo is given by: 
\begin{equation}
V_{i}^{B}=\ln O_{sj_{i}^{B}}+W_{sj_{i}^{B}}^{\prime }\beta +Z_{i}^{\prime
}\delta _{sj_{i}^{B}}+R_{j_{i}^{B}}^{\prime }\alpha +\xi
_{j_{i}^{B}}+\epsilon _{isj_{i}^{B}}
\end{equation}%
In contrast, if the identical agent were exposed to a fully transparent
information regime that eliminated cognitive frictions (ensuring $\gamma =1$%
), the counterfactual rational choice would be: 
\begin{equation}
j_{i}^{R}=\arg \max_{j\in \{1,\dots ,J\}}\left\{ \ln O_{sj}+W_{sj}^{\prime
}\beta +Z_{i}^{\prime }\delta _{sj}+R_{j}^{\prime }\alpha +\xi _{j}+\epsilon
_{isj}\right\}
\end{equation}%
yielding an experienced welfare level of $V_{i}^{R}$. By revealed preference
under objective probabilities, it follows that $V_{i}^{B}\leq V_{i}^{R}$.
The difference $V_{i}^{R}-V_{i}^{B}\geq 0$ represents a pure behavioral
deadweight loss. Subjective probability distortions thus increase the
inefficiency already arising from physical congestion.

\subsection{Semiparametric Identification of the Behavioral Parameter}

We now establish that the structural behavioral parameter $\gamma $ is
point-identified alongside the taste parameters without requiring parametric
assumptions on the distribution of preference heterogeneity. Let alternative 
$J$ be the baseline reference option. Under standard multinomial index
inversion approaches (e.g., Ahn et al., 2018; Khan et al., 2021), the
conditional choice probabilities directly identify the systematic utility
contrasts: 
\begin{equation}
t_{sij}=A_{sj}+Z_{i}^{\prime }B_{sj}
\end{equation}%
where the intercept $A_{sj}$ captures the market-level common components
relative to option $J$: 
\begin{equation}
A_{sj}=(W_{sj}-W_{sJ})^{\prime }\beta +(\tilde{\xi}_{j}-\tilde{\xi}%
_{J})+g(O_{sj};\gamma )-g(O_{sJ};\gamma )
\end{equation}%
and $\tilde{\xi}_{j}\equiv R_{j}^{\prime }\alpha +\xi _{j}$ represents the
composite time-invariant college component. The within-subject slope
contrast is identified as $B_{sj}=\delta _{sj}-\delta _{sJ}$.

To isolate the behavioral parameter $\gamma$ from the unobserved realized
option fixed effects $\tilde{\xi}_{j}$, we exploit the panel-like structure
across repeated markets (subjects of study $s$ and $r$). For any pair of
markets $(s, r)$, let the cross-subject difference in observed attributes
be: 
\begin{equation}
\Delta_{sr} W_{j} \equiv (W_{sj} - W_{sJ}) - (W_{rj} - W_{rJ})
\end{equation}
and let the cross-subject difference in the transformed objective success
probabilities be: 
\begin{equation}
\Delta_{sr} G_{j}(\gamma) \equiv [g(O_{sj}; \gamma) - g(O_{sJ}; \gamma)] -
[g(O_{rj}; \gamma) - g(O_{rJ}; \gamma)]
\end{equation}
Differencing the identified first-stage intercepts $A_{sj}$ across subjects
cleanly eliminates the common unobserved college fixed effects: 
\begin{equation}
A_{sj} - A_{rj} = \Delta_{sr} W_{j}^{\prime }\beta + \Delta_{sr}
G_{j}(\gamma)
\end{equation}

Let $X$ denote the stacked matrix whose rows correspond to the covariate
differences $\Delta_{sr} W_{j}^{\prime }$, and let $G(\gamma)$ stack the
corresponding probability weighting contrasts $\Delta_{sr} G_{j}(\gamma)$.
The complete system can be written compactly as $\Delta A = X\beta +
G(\gamma)$.

\begin{proposition}
Suppose that the first-stage inversion identifies the systematic utility
contrasts $\{A_{sj}, B_{sj}\}$, the coefficient on the transformation
function $g(O_{sj}; \gamma)$ is normalized to one, and for any $(\beta,
\gamma) \neq (\beta_0, \gamma_0)$ there exists at least one triplet $(j, s,
r)$ such that: 
\begin{equation}
\Delta_{sr} W_{j}^{\prime }(\beta - \beta_0) + \Delta_{sr} G_{j}(\gamma) -
\Delta_{sr} G_{j}(\gamma_0) \neq 0
\end{equation}
If the matrix $X$ has full column rank and we define the residual maker
matrix $M_{X} = I - X(X^{\prime -1}X^{\prime }$, then a sufficient condition
for the global identification of the behavioral parameter $\gamma$ is: 
\begin{equation}
M_{X}G(\gamma) \neq M_{X}G(\gamma_0) \quad \forall \quad \gamma \neq \gamma_0
\end{equation}
Furthermore, a sufficient condition for local identification around the true
parameter value $\gamma_0$ is: 
\begin{equation}
\text{rank} 
\begin{bmatrix}
X & G_{\gamma}(\gamma_0)%
\end{bmatrix}
= \text{dim}(\beta) + 1
\end{equation}
where $G_{\gamma}(\gamma_0)$ stacks the analytical derivatives: 
\begin{equation}
\frac{\partial g(O_{sj}; \gamma_0)}{\partial \gamma} = -(-\ln
O_{sj})^{\gamma_0} \ln(-\ln O_{sj})
\end{equation}
\end{proposition}

\begin{proof}
The first-stage inversion uniquely pinpoints the vector of intercept
differences $\Delta A$ from the conditional choice data. Given the rank
condition on $X$ and the global uniqueness condition, the mapping from $%
(\beta, \gamma)$ to $\Delta A$ is injective. Specifically, pre-multiplying
the system by $M_{X}$ yields $M_{X}\Delta A = M_{X}G(\gamma)$. Because $%
M_{X}G(\gamma)$ is strictly injective under the assumption, $\gamma$ is
uniquely determined. Once $\gamma$ is recovered, $\beta$ is recovered via
OLS as $\beta = (X^{\prime -1}X^{\prime }(\Delta A - G(\gamma))$. Under the
maintained orthogonality or correlated random effects restrictions on $%
\tilde{\xi}_{j}$, the preference parameters $\alpha$ and relative
unobservables $\xi_{j} - \xi_{J}$ are subsequently identified.
\end{proof}

The intuition behind this identification result is that variation in
objective success rates across distinct academic subjects breaks the
collinearity with fixed college traits. Because $g(O_{sj};\gamma )$ is
non-linear in $\gamma $, variation in objective capacities shifts the
relative transformed capacities across subjects in a manner that cannot be
absorbed by linear adjustments to the structural coefficients $\beta $. This
allows the data to separate standard preference parameters from those that
characterize cognitive probability distortions.

\subsection{The Effect of Information Campaigns}

From a public policy perspective, knowledge of $\gamma $ allows the
researcher to simulate the effects of targeted informational campaigns.
Evidence from college choice shows that high-achieving students from
low-income backgrounds can be misinformed about their opportunities, and
that targeted information provision can improve their knowledge and change
their decisions (cf. Hoxby and Turner, 2015). By estimating
demographic-specific distortion parameters $\gamma _{m}\neq \gamma _{f}$
where $m,f$ \ represent different demographic groups, our framework can
simulate a counterfactual intervention that enhances informational
transparency. Such policies reduce subjective misperceptions ($\gamma
\rightarrow 1$) without expensive and/or distortive interventions like
modifying prices, capacities, or choice sets. The welfare gain of this
informational campaign for individual $i$, expressed as compensating
variation in terms of the numeraire price/rent coefficient $\alpha $, is
single-valued and point-identified: 
\begin{equation}
\Delta T_{i}=\frac{V_{i}^{R}-V_{i}^{B}}{\alpha }
\end{equation}

Evaluating this distribution exposes the precise equity implications of
informational campaigns. If informational frictions are heavily concentrated
among specific demographic subgroups, public outreach campaigns can generate
substantial, Pareto-improving welfare adjustments that reduce systemic
inequalities without requiring market-distorting price interventions. In
fact, the University of Cambridge runs a intensive outreach campaign in
areas that have historically sent very few students to Cambridge.\footnote{%
cf. www.undergraduate.study.cam.ac.uk/find-out-more/widening-participation}
The values of $\gamma $ across different groups or areas can inform
efficient targeting of such interventions.

\section{Future Research}

We conclude this paper with a suggestion for future work. A potential
intervention in the Cambridge admission case is to change the existing
decentralized system to a centralized one similar to the one used to match
applicants to primary and secondary public schools in the US. This follows
the seminal work of Gale, Shapley and Roth, where one elicits preferences
for schools from applicants and those for students from schools, and uses a
deferred acceptance type centralized algorithm to construct the match. It
would be interesting to explore if the welfare effect of a switch to such an
algorithmic procedure can be identified using solely data from the status
quo situation where admissions are decentralized. Note, however, that
Cambridge (and Oxford) colleges do not have heterogeneous preferences across
applicants and follow the university-wide admission criteria and objectives
stipulated by the subject's faculty. This is analogous to modes of transport
not having heterogeneous preferences across commuters. So, one may need to
modify the goal of such centralized matching, e.g. balance the objectives of
admitting the academically most promising students while respecting their
preferences for colleges. Indeed, our analysis in this paper provides the
decentralized benchmark, i.e. free individual choice under capacity
constraints, for comparison to alternative interventions, including
switching to Gale-Shapley type preference-based centralized matching or
expansion of capacity -- questions that we leave for future research.

\pagebreak

\begin{center}
\textbf{References}
\end{center}

\begin{enumerate}
\item Ahn, H., Ichimura, H., Powell, J.L. and Ruud, P.A., 2018. Simple
estimators for invertible index models. Journal of Business \& Economic
Statistics, 36(1), pp.1-10.

\item Berry, S., Gandhi, A. and Haile, P., 2013. Connected substitutes and
invertibility of demand. Econometrica, 81(5), pp.2087-2111.

\item Berry, S.T. and Haile, P.A., 2024. Nonparametric identification of
differentiated products demand using micro data. Econometrica, 92(4),
pp.1135-1162.

\item Bhattacharya, D., 2015. Nonparametric welfare analysis for discrete
choice. Econometrica, 83(2), pp.617-649.

\item Bhattacharya, D., 2018. Empirical welfare analysis for discrete
choice: Some general results. Quantitative Economics, 9(2), pp.571-615.

\item Bhattacharya, D., Dupas, P. and Kanaya, S., 2024. Demand and welfare
analysis in discrete choice models with social interactions. Review of
Economic Studies, 91(2), pp.748-784.

\item Bhattacharya, D., Kanaya, S. and Stevens, M., 2017. Are university
admissions academically fair?. Review of Economics and Statistics, 99(3),
pp.449-464.

\item Bhattacharya, D. and Shvets, J., 2024. Inferring trade-offs in
university admissions: Evidence from Cambridge. Journal of Political
Economy, 132(11), pp.3737-3784.

\item Brock, W.A. and Durlauf, S.N., 2001. Discrete choice with social
interactions. The Review of Economic Studies, 68(2), pp.235-260.

\item Brock, W.A. and Durlauf, S.N., 2002. A multinomial-choice model of
neighborhood effects. American Economic Review, 92(2), pp.298-303.

\item Brock, W.A. and Durlauf, S.N., 2007. Identification of binary choice
models with social interactions. Journal of Econometrics, 140(1), pp.52-75.

\item Brock, W. and Durlauf, S., 2011. Identification of social
interactions', in (J. Benhabib, A. Bisin and MO Jackson, eds.), Handbook of
Social Economics, vol. 1.

\item Chetty, R., Looney, A. and Kroft, K., 2009. Salience and taxation:
Theory and evidence. American economic review, 99(4), pp.1145-1177.

\item Hausman, J.A. and Newey, W.K., 2016. Individual heterogeneity and
average welfare. Econometrica, 84(3), pp.1225-1248.

\item Hausman, J.A. and Newey, W.K., 2017. Nonparametric welfare analysis.
Annual Review of Economics, 9(1), pp.521-546.

\item Hotz, V.J. and Miller, R.A., 1993. Conditional choice probabilities
and the estimation of dynamic models. The Review of Economic Studies, 60(3),
pp.497-529.

\item Hoxby, Caroline M., and Sarah Turner. 2015. What High-Achieving
Low-Income Students Know about College. American Economic Review, 105(5):
514--517.

\item Khan, S., Ouyang, F. and Tamer, E., 2021. Inference on semiparametric
multinomial response models. Quantitative Economics, 12(3), pp.743-777.

\item Lewbel, A., 2001. Demand Systems with and without Errors. American
Economic Review, 91(3), pp.611-618.

\item McFadden, D., 1972. Conditional logit analysis of qualitative choice
behavior. https://escholarship.org/uc/item/61s3q2xr

\item Milchtaich, I., 2000. Generic uniqueness of equilibrium in large
crowding games. Mathematics of Operations Research, 25(3), pp.349-364.

\item Prelec, D., 1998. The probability weighting function. Econometrica,
pp.497-527.

\item Stantcheva, S., 2021. Understanding tax policy: How do people reason?.
The Quarterly Journal of Economics, 136(4), pp.2309-2369.
\end{enumerate}

\pagebreak

\begin{center}
{\LARGE Appendix}
\end{center}

\textbf{A. Relative congestion dominance holds under Cambridge pool:}

Suppress the subject index and let $\pi =(\pi _{1},\ldots ,\pi _{J})$ denote
the vector of initial application shares, where $\pi _{j}\geq 0$ and $%
\sum_{j=1}^{J}\pi _{j}=1$. Let $U_{ij}(\pi )$ be applicant $i$'s expected
utility from applying initially to college $j$, including any subsequent
admission through the pool. Normalize the payoff from rejection to zero.

For two share vectors $\pi $ and $\pi ^{\prime }$, define 
\begin{equation}
A(\pi ,\pi ^{\prime })=\{j:\pi _{j}\geq \pi _{j}^{\prime }\}.
\end{equation}%
The relative congestion dominance part of Assumption 1 requires that, for
every applicant $i$, $a\in A(\pi ,\pi ^{\prime })$ and $b\notin A(\pi ,\pi
^{\prime })$, 
\begin{equation}
U_{ia}(\pi )-U_{ib}(\pi )\leq U_{ia}(\pi ^{\prime })-U_{ib}(\pi ^{\prime }).
\label{dominance}
\end{equation}

\begin{proposition}[Pooling with college-specific continuation values]
For each applicant $i$, suppose that:

\begin{enumerate}
\item Direct admission following an initial application to college $j$
occurs with probability $p_{ij}(\pi_j)$, which is weakly decreasing in $%
\pi_j $ and does not depend on other application shares.

\item Direct admission to college $j$ yields payoff $H_{ij}$, independent of
application shares.

\item Conditional on not receiving direct admission, the expected payoff
from the pool, including the possibility of rejection, is $V_{ij}$. This
continuation value may depend on the initial application college $j$ but is
independent of the share vector. No equality across initial colleges is
required.

\item $H_{ij}\geq V_{ij}$ for every college $j$.
\end{enumerate}

Then expected utilities satisfy (\ref{dominance}). If applicants also
correctly anticipate equilibrium application shares, Assumption 1 holds.
\end{proposition}

\begin{proof}
Expected utility from applying initially to college $j$ is 
\begin{eqnarray}
U_{ij}(\pi ) &=&p_{ij}(\pi _{j})H_{ij}+[1-p_{ij}(\pi _{j})]V_{ij}  \notag \\
&=&V_{ij}+p_{ij}(\pi _{j})(H_{ij}-V_{ij}).  \label{constantvalue}
\end{eqnarray}%
Since $H_{ij}-V_{ij}\geq 0$, expected utility is weakly decreasing in $\pi
_{j}$ and independent of all other shares. Take any two share vectors $\pi
,\pi ^{\prime }$ and let $A=A(\pi ,\pi ^{\prime })$. For $a\in A$ and $%
b\notin A$, therefore, 
\begin{equation}
U_{ia}(\pi )\leq U_{ia}(\pi ^{\prime }),\qquad U_{ib}(\pi )\geq U_{ib}(\pi
^{\prime }).
\end{equation}%
Subtracting yields (\ref{dominance}).
\end{proof}

The continuation value may accommodate heterogeneous preferences over both
initial and destination colleges. Let $q_{ijk}$ be the probability of
eventual admission to college $k$, conditional on direct non-admission
following an initial application to college $j$. Then 
\begin{equation}
V_{ij}=\sum_{k=1}^{J}q_{ijk}H_{ik}, \qquad q_{ijk}\geq 0,\qquad
\sum_{k=1}^{J}q_{ijk}\leq 1.
\end{equation}
The remaining probability corresponds to rejection, including failure to
enter the pool. The conditional lottery may differ across initial colleges.
An invariant lottery for each initial college is sufficient; more generally,
its probabilities may respond to congestion provided that its expected
payoff remains the same $V_{ij}$ for that initial college.

This result reinforces the unique equilibrium result of Theorem 1 even in
the specific case of Cambridge applications that involves a pool.

\bigskip

\textbf{B. Example showing unique fixed point and invertibility are
unrelated:}

The following examples show that invertibility of $G\left( \cdot \right) $
neither implies nor is implied by uniqueness of its fixed point. First,
consider the function $G\left( \cdot \right) :\left[ 0,1\right] \rightarrow
\lbrack 0,1]$ given by%
\begin{equation*}
G\left( x\right) =\left\{ 
\begin{array}{l}
1.5x\text{ if }0\leq x\leq 0.25 \\ 
0.5x+0.25\text{ if }0.25\leq x\leq 0.75 \\ 
1.5x-0.5\text{ if }0.75\leq x\leq 1\text{.}%
\end{array}%
\right.
\end{equation*}%
The function is plotted below in blue, and the red straight line is the 45
degrees line.\begin{center}\includegraphics[width=0.58\linewidth]{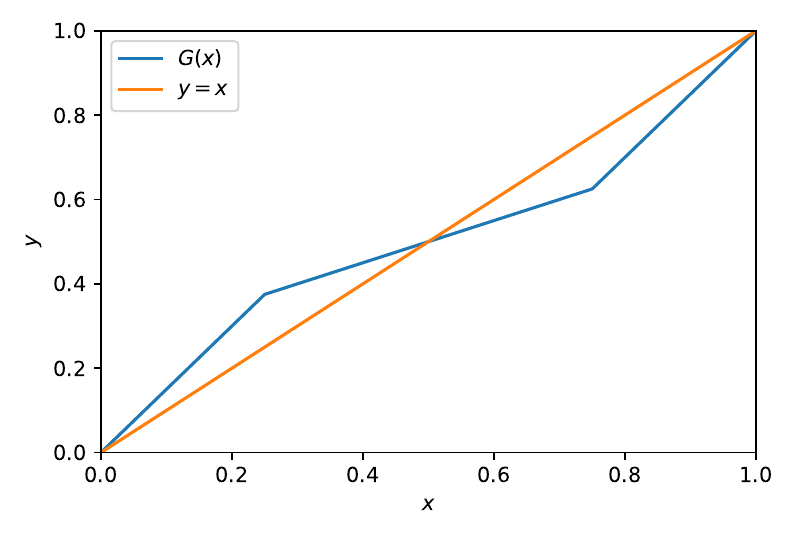}\end{center}

\clearpage
\section*{Empirical Tables and Figures}
\addcontentsline{toc}{section}{Empirical Tables and Figures}
\includepdf[pages=-,pagecommand={},fitpaper=false]{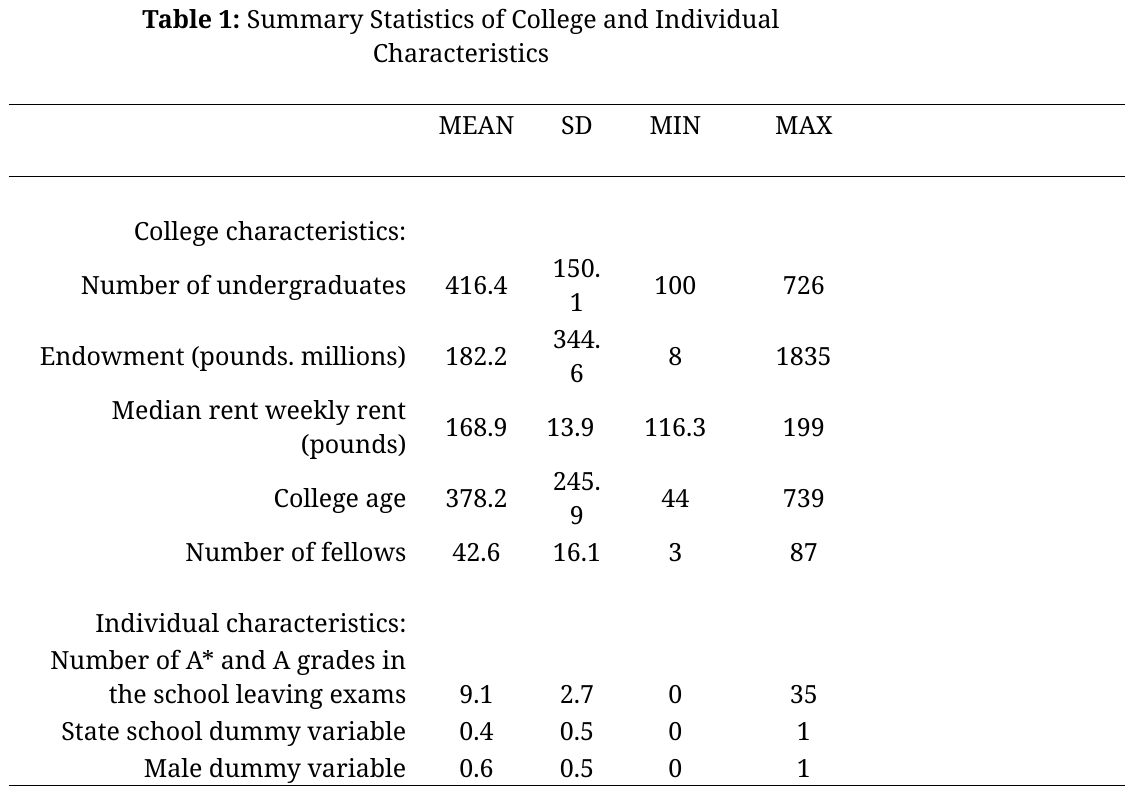}
\end{document}